\documentclass[11pt,prd,letterpaper,tightenlines,superscriptaddress]{revtex4}
\usepackage{amsmath,amssymb,amsfonts,graphicx,dcolumn}
\usepackage{epstopdf}
\usepackage{color}
\usepackage[utf8]{inputenc}
\usepackage{amsthm}
\usepackage{fontenc}
\usepackage{slashed}
\usepackage[dvips]{hyperref}
\usepackage{bm}
\usepackage{graphics}
\RequirePackage{slashed}
\usepackage{cancel}
\usepackage[normalem]{ulem}

\usepackage{array,booktabs}
\newcolumntype{C}{>{$\displaystyle}c<{$}}

\begin{document}


\title{$Z$ plus jets production via double parton scattering in $pA$ collisions at the LHC}

\author{Boris Blok}
\email{blok@physics.technion.ac.il}
\affiliation{ Department of Physics, Technion, Israel Institute of Technology, Haifa, 32000 Israel}
\author{Federico Alberto Ceccopieri}
\email{federico.ceccopieri@hotmail.it}	
\affiliation{ Department of Physics, Technion, Israel Institute of Technology, Haifa, 32000 Israel}
\affiliation{IFPA, Université de Li\`ege, B4000, Li\`ege, Belgium}

\begin{abstract}
We present results on $Zjj$ production via double parton scattering in $pA$ collisions at the LHC. 
We perform the analysis at leading and next-leading order accuracy with different sets of cuts on jet transverse momenta
and accounting for the single parton scattering background.
By exploiting the experimental capability to measure the centrality dependence of the cross section,  we discuss the  feasibility of DPS observation 
in already collected data at the LHC and in future runs. 
\end{abstract}

\maketitle

\section{Introduction}
\label{Sec:Intro}
\par The study of multiple parton interaction (MPI) and in particular of hard double parton scattering (DPS) reactions in  $pA$ collisions is important for our understanding of  MPI in $pp$ collisions. Significant progresses were achieved in study of double parton scattering in proton-nucleus collisions for a variety of final states~\cite{Helenius:2019uge,sde,dEnterria:2012jam,Cattaruzza:2004qb,Calucci:2013pza,Cazaroto:2016nmu,Huayra:2019iun,Shao:2020acd}
and implemented in PYTHIA Monte Carlo simulation~\cite{Fedkevych:2019ofc}. 

The theory of DPS in $pA$ collisions was developed in~\cite{Strikman:2001gz,BSW}, where it was shown that there are two DPS contributions at work in such a case. 
First, there is the so-called DPS1 contribution, in which two partons from the incoming nucleon interact with two partons in the target nucleon in the nucleus and which is formally identical to DPS in $pp$ collisions \cite{TreleaniPaver82,mufti,stirling,BDFS1,Diehl,stirling1,BDFS2,Diehl2,BDFS3,BDFS4,Diehl:2017kgu,Manohar:2012jr}. Then there is a new type of contribution, 
which we refer to as DPS2, in which 
two partons from the incoming nucleon interact with two partons  each of them belonging to the distinct nucleons in the target nucleus located at the same impact parameter. 
\par Recently a new method was suggested~\cite{Alvioli:2019kcy} which could allow the observation of DPS2 in $pA$ collisions. It was pointed out that the DPS2 has a different dependence on impact parameter than single parton scattering (SPS) and DPS1 contributions.
Namely while the latter contributions are  proportional to  the nuclear thickness function $T(B)$, $B$ being the $pA$ impact parameter, the  DPS2 contribution is proportional to the square of $T(B)$. Therefore the cross section for producing a given final state can be schematically written as:
\begin{equation}
\frac{d^2\sigma_{pA}}{d^2B}= 
\Big(\sigma^{LT}_{pA}+\sigma^{DPS1}_{pA}\Big) \frac{T(B)}{A} +\sigma_{pA}^{DPS2} \frac{T^2(B)}{\int d^2 B \, T^2(B)}\,,
\label{e1}
\end{equation}
where $T(B)$ is normalized to the atomic number $A$ of the nucleus. 
This observation gives the possibility to distinguish the DPS2
contribution in $pA$ collisions from both the leading twist (LT) SPS and DPS1 contributions that are instead linear in $T(B)$.
This strategy has been adopted in our  recent papers  where we have analyzed the associated production of electroweak $W$ boson and jets~\cite{Blok:2019fgg} and multijet production~\cite{Blok:2020oce}  via DPS in $pA$ collisions.
There we have shown that, exploiting the experimental capability of measuring the centrality dependence of the cross section~\cite{25,30,35}, one can 
separate the  DPS2 mechanism exploiting its different dependence on $T(B)$, as it appears from Eq.~(\ref{e1}).
We found that the procedure can be successfully carried on for those final states by using the data already recorded in 2016 $pA$ runs at the LHC.
\par In this study  we shall extend those results to the $Zjj$ final state. We will show that, even in this case, by applying the very same technique, one can separate the DPS2 contribution from the DPS1+SPS background, despite the lower event rate associated to $Z$ production, as compared, for example, to the ones for final states analysed in Refs.~\cite{Blok:2019fgg,Blok:2020oce}. 
\par We present our results at leading order (LO) and 
next-to-leading order (NLO) accuracy.
In the former approximation we find that, by using symmetric cuts on jet transverse momenta,  it will be possible to observe the DPS2 contribution in $pA$ data already collected at LHC.
To NLO accuracy we were able to study only 
the case of asymmetric cuts, \textsl{i.e.} $p_{1T}^{cut}-p_{2T}^{cut}\ge 10 $ GeV, for the difference in the transverse momentum cuts of the leading and subleading jet. We adopted this prescription in order to tackle  reliability issues  inherent to the dijet NLO calculation as the difference of jet transverse momentum threshold is lowered. In the limit of vanishing transverse momentum difference, \textsl{i.e.} in the symmetric cut limit,  
the predictivity of the theory is recovered by performing an all order soft gluon resummation  which is, however, beyond the scope of the current paper.
The analysis within asymmetric cuts shows that NLO corrections lead to a slightly stronger DPS2 signal with respect to LO ones.  However the statistical significance of the DPS2 signal decreases 
as a result of the reduced dijet rates obtained with asymmetric cuts choice  for which we were able to determine NLO corrections. In such a case  we may need higher statistics  for detailed analysis of $Zjj$ final state, although the signal can be appreciated already within the available $pA$ with the lowest jet transverse momentum threshold.

\par The paper is organized as follows. In Section \ref{Sec:Calculation}
we briefly review the theoretical formalism at the base of our calculations. 
In Section \ref{Sec:LO} we present our results at leading order  accuracy with
symmetric cuts on the jet transverse momenta. In Section \ref{Sec:NLO}
we present our results at leading and next-to-leading order accuracy with
asymmetric cuts on the jet transverse momenta. We summarize our findings
in Section \ref{Sec:Conclusions}.

\section{Calculation}
\label{Sec:Calculation}
In this paper we consider the production of $Z$ boson plus dijet in proton-lead collisions:
\begin{equation}
p P_b \rightarrow Z + 2jets + X\,, \nonumber
\end{equation}
where the $Z$ decays leptonically and at least two jets are found in the final state. The corresponding DPS cross section (with $C=Z$ and $D=jj$) is written
to leading order accuracy as~\cite{Strikman:2001gz,BSW,Blok:2019fgg,Blok:2020oce}: 
\begin{multline}
\frac{d \sigma^{CD}_{DPS}}{d\Omega_1 d\Omega_2} =  
\sum_{i,j,k,l} 
\sum_{N=p,n} \sigma_{eff}^{-1}
 f^{i}_{p}(x_1) f^{j}_{p}(x_2)
f^{k}_{N}(x_3) f^{l}_{N}(x_4) 
\frac{d \hat \sigma_{ik}^C}{d\Omega_C} \frac{d \hat \sigma_{jl}^D}{d\Omega_D} \int d^2 B \, T_N(B)+\\
+ \sum_{i,j,k,l} \sum_{N_3, N_4=p,n}
f^{i}_{p}(x_1) f^{j}_{p}(x_2)
f^{k}_{N_3}(x_3) f^{l}_{N_4}(x_4) 
\frac{d \hat \sigma_{ik}^C}{d\Omega_C} \frac{d \hat \sigma_{jl}^D}{d\Omega_D}\,
\int d^2 B \, T_{N_3}(B) T_{N_4}(B)\,. 
\label{due}
\end{multline}

\par   The nuclear thickness function $T_N(B)$ appearing in Eq.~(\ref{due}), is obtained integrating the proton and neutron densities $\rho_0^{(p,n)}(B,z)$ in the nucleus over the longitudinal component $z$.
Following Ref.~\cite{Alvioli:2018jls,Blok:2019fgg,Blok:2020oce}, for the ${}^{208} P_b$ nucleus, the density of proton and neutron is described by a Wood-Saxon distribution
whose parameters are fixed according to the analyses of Refs.~\cite{Warda:2010qa,Tarbert:2013jze}.

The first term in Eq.~(\ref{due}) corresponds to the DPS1 mechanism, linear in the nuclear thickness function $T_A$. It is calculated by assuming $\sigma_{eff}=$18 mb, the average of experimental extracted value for similar final states~\cite{ATLAS_WJJ,CMS_WJJ} in DPS analyses in $pp$ collisions. 
The second term corresponds to the DPS2 mechanism and it is  quadratic in $T_A$
~\cite{Strikman:2001gz}.
Double distributions appearing in the DPS1 and DPS2 contributions are evaluated in mean field approximation, \textsl{i.e.} we assume that they can be written as a product of single parton distributions, as already assumed in Eq.~(\ref{due}). 
Quite importantly, the DPS2 term on the second line involves one double distribution integrated over the partonic relative 
interdistance and therefore is free of the inherent uncertainty introduced by $\sigma_{eff}$ which affects the DPS1 term.
In the same equation $d\hat{\sigma}/d\Omega$ stand
for partonic cross sections differential in the relevant set of variables $\Omega$. 

The production of $Z$ boson in proton-lead collisions has been measured 
at 5.02 TeV in Ref.~\cite{Khachatryan:2015pzs} and it has been found to scale with the atomic number $A$ of the colliding nucleus to very good approximation. The associated production of jets associated with $Z$ boson in $pp$ at 7 TeV has been measured in Ref.~\cite{Aad:2013ysa}.

From those analyses we use the following cuts and settings.
We set the per-nucleon centre-of-mass energy $\sqrt{s_{pN}}$ = 8.16 TeV, with proton 
energy $E_p = 6.5$ TeV and nucleon energy $E_{N} = 2.56$ TeV.
The  proton-nucleon centre-of-mass is boosted with respect to the
laboratory frame by $\Delta y = 1/2 \, \ln E_p/E_N$ = 0.465 in the proton direction, assumed to be at positive rapidity. Therefore the rapidity shift reads $y_{CM}=y_{lab}-\Delta y$. In all calculations, we consider proton-nucleon centre-of-mass rapidities.
For the $Z$ kinematics we require the lepton rapidities to lie in $|\eta_l^{lab}|<2.4$, 
their transverse momentum to be $p_T^l>20$ GeV. We also require the dilepton invariant mass to be in the range  66 GeV$< m^{ll} <$ 116 GeV and the jet-lepton distance to be $\Delta R^{lj}>$0.5. The $Z$ fiducial cross sections take into account its decay in electron and muon pair. 
Jets are clustered at parton level according to anti-$k_t$ jet algorithm with jet radius $R=0.7$. The dijet kinematics is restricted to jet rapidities $|\eta_j^{lab}|<4.4$. 
\par The LO results presented in 
Section \ref{Sec:LO} are obtained with symmetric cuts on jet transverse momenta, 
while LO and NLO predictions presented in 
Section \ref{Sec:NLO} are obtained with asymmetric ones.
Details on additional settings on the simulations in these 
two cases can be found in the corresponding Sections. 

\section{Leading order results with symmetric pt-cuts}
\label{Sec:LO}
The $Z$ and $Zjj$ SPS cross sections have been calculated to leading order accuracy with
\texttt{MCFM}~\cite{MCFM} by using \texttt{CTEQ6L1}~\cite{Pumplin:2002vw} free proton parton distributions supplemented with nuclear corrections factors  from Ref.~\cite{EKS98}. For both processes the renormalization, $\mu_R$, and factorization scale, $\mu_F$, are both fixed to the $Z$-boson mass, $m_Z$, the only available option in the code.  
The dijet cross sections have been calculated at leading order accuracy by using  \texttt{ALPGEN}~\cite{Mangano:2002ea} generator 
with the choice $\mu_R=\mu_F=\sqrt{p_{T,1}^2+p_{T,2}^2}$, being 
$p_{T,i}$ the transverse momentum of jet $i$.  
In this case nuclear effects have been neglected, since
our LO estimates indicate that they reduce the cross section by a few percent.
We consider symmetric cuts on jet transverse momenta, $p_T^j$, in three scenarios 
in which both of them are required to have 
$p_T^{j_1,j_2}>$20, $p_T^{j_1,j_2}>$25 and $p_T^{j_1,j_2}>$30 GeV, respectively. 


\setlength{\extrarowheight}{0.2cm}
\begin{table}[t]
\begin{center}
\begin{tabular}{cccccccc}  \hline  \hline 
 & \hspace{0.4cm} DPS1 \hspace{0.4cm} & \hspace{0.4cm} DPS2 \hspace{0.4cm}  
 & \hspace{0.4cm} SPS \hspace{0.4cm} & \hspace{0.4cm} Sum \hspace{0.4cm} 
 & \hspace{0.4cm} $\sigma(Zjj)/\sigma(Z)$ \hspace{0.4cm} &  $f_{DPS1}$  &  \hspace{0.4cm}  $f_{DPS2}$ \\  
$Zjj$ &  [pb] & [pb] & [pb] & [pb] &  & & \\ \hline
$p_T^{j_1,j_2}>20,20$ GeV &  2971 &  7814 & 15940 & 26725 & 0.166  &  0.111 &   \hspace{0.4cm}   0.292  \\
$p_T^{j_1,j_2}>25,25$ GeV &  1270 &  3341 & 11024 & 15636  & 0.097  &  0.081 &   \hspace{0.4cm}   0.213  \\   
$p_T^{j_1,j_2}>30,30$ GeV &  621  &  1632 & 8030 & 10283   & 0.064  &  0.060 &   \hspace{0.4cm}   0.158   \\
\hline
\end{tabular}
\caption{\textsl{Leading order predictions for $Zjj$ DPS and SPS cross sections in $pA$ collisions in fiducial phase space, for symmetric cuts on jets transverse momenta. The last three columns display the cross sections ratios as explained in the text.}}
\label{Zjj:LOcs}
\end{center}
\end{table}

\begin{figure}[t]
\includegraphics[scale=0.6]{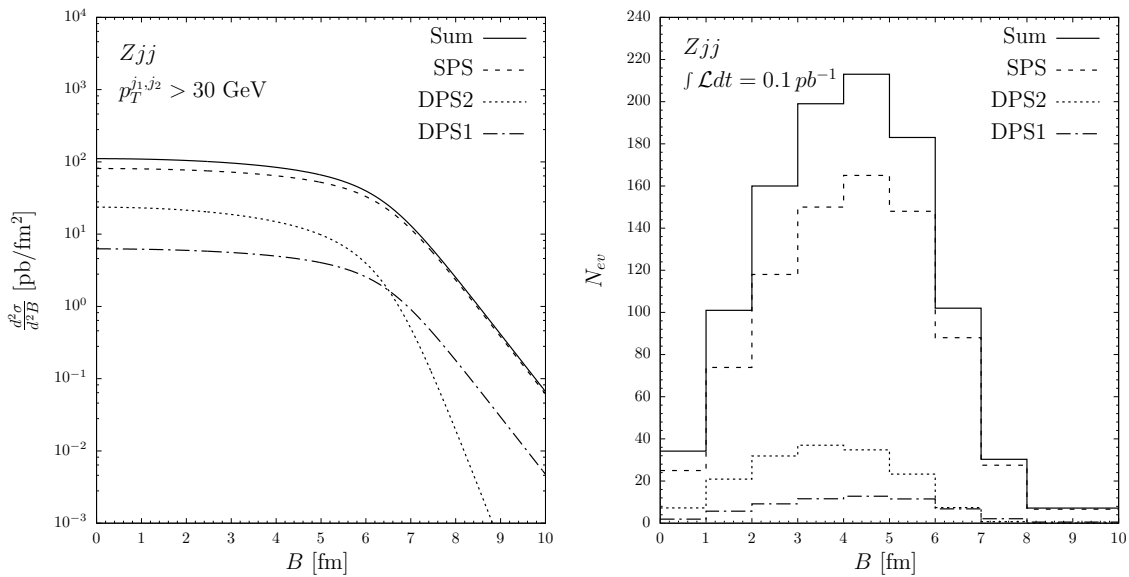}
\caption{\textsl{Left panel : various contributions to the $Zjj$ cross sections as a function of $B$. Right panel: expected number of events integrated in bins of $B$ for  $\int \mathcal{L} \mbox{dt}= 0.1 \, \mbox{pb}^{-1}$.
Both plots are obtained with $p^j_T>30$ GeV.}}
\label{plot:NevZjj}
\end{figure}

\setlength{\extrarowheight}{0.2cm}
\begin{table}[t]
\begin{center}
\begin{tabular}{cccccc}  \hline  \hline 
 && \hspace{0.4cm}  $p_T^{j_1,j_2}>20$ GeV \hspace{0.4cm}  & \hspace{0.4cm}  $p_T^{j_1,j_2}>25$ GeV \hspace{0.4cm}  & \hspace{0.4cm}  $p_T^{j_1,j_2}>30$ GeV \hspace{0.4cm}  & \\ \hline
 $T_{min}$  & $T_{max}$ &  $N^{Zjj}_{ev}$ &    $N^{Zjj}_{ev}$ &   $N^{Zjj}_{ev}$ &  $N^{Z}_{ev}$ \\ \hline
  0.0 &  0.9 &  356   & 222   &  152  &  2657 \\
  0.9 &  1.7 &  1075  & 632   &  417  &  6605 \\
  1.7 &  2.1 &  1241  & 710   &  459  &  6872 \\ \hline
\end{tabular}
\caption{\textsl{Number of expected $Zjj$ and $Z$ events assuming  $\int \mathcal{L} \mbox{dt}= 0.1 \, \mbox{pb}^{-1}$,  integrated in bins of $T_A$
with $Z$ decaying into opposite sign electrons and muons. The Number of $Zjj$ is reported for three different symmetric cuts on jet transverse momenta. Cross sections are evaluated to leading order accuracy. }}
\label{Nev_zjj_LO_sym}
\end{center}
\end{table}

We report in Tab.~(\ref{Zjj:LOcs}) 
the various DPS and SPS contributions to the $Zjj$ fiducial cross section for three different transverse momentum cuts on the jets. In the last three columns
we report the ratio of the total $Zjj$ (SPS+DPS) over inclusive $Z$ cross section and the relative fractions of both DPS contributions over the total $Zjj$ cross section, \textsl{i.e.}  $f_{DPS1}=\sigma^{DPS1}(Zjj)/\sigma^{Sum}(Zjj)$ and $f_{DPS2}=\sigma^{DPS2}(Zjj)/\sigma^{Sum}(Zjj)$. From these  ratios one can appreciate the increase of the DPS fractions as the cuts on jet transverse momenta are lowered and that, on average, the DPS2 contribution is nearly three times larger than DPS1. 

\begin{figure}[t]
\includegraphics[scale=0.6]{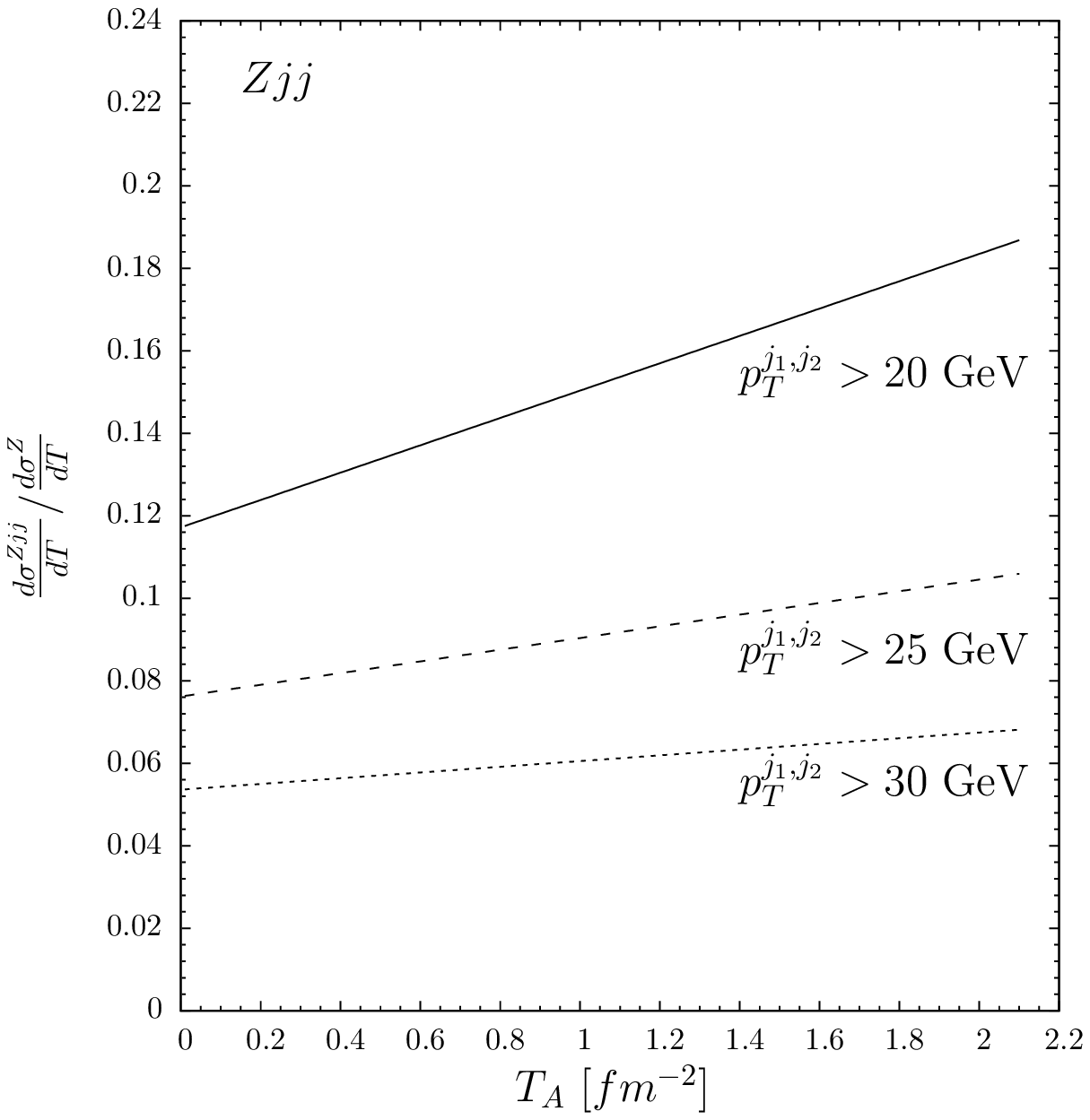}
\includegraphics[scale=0.6]{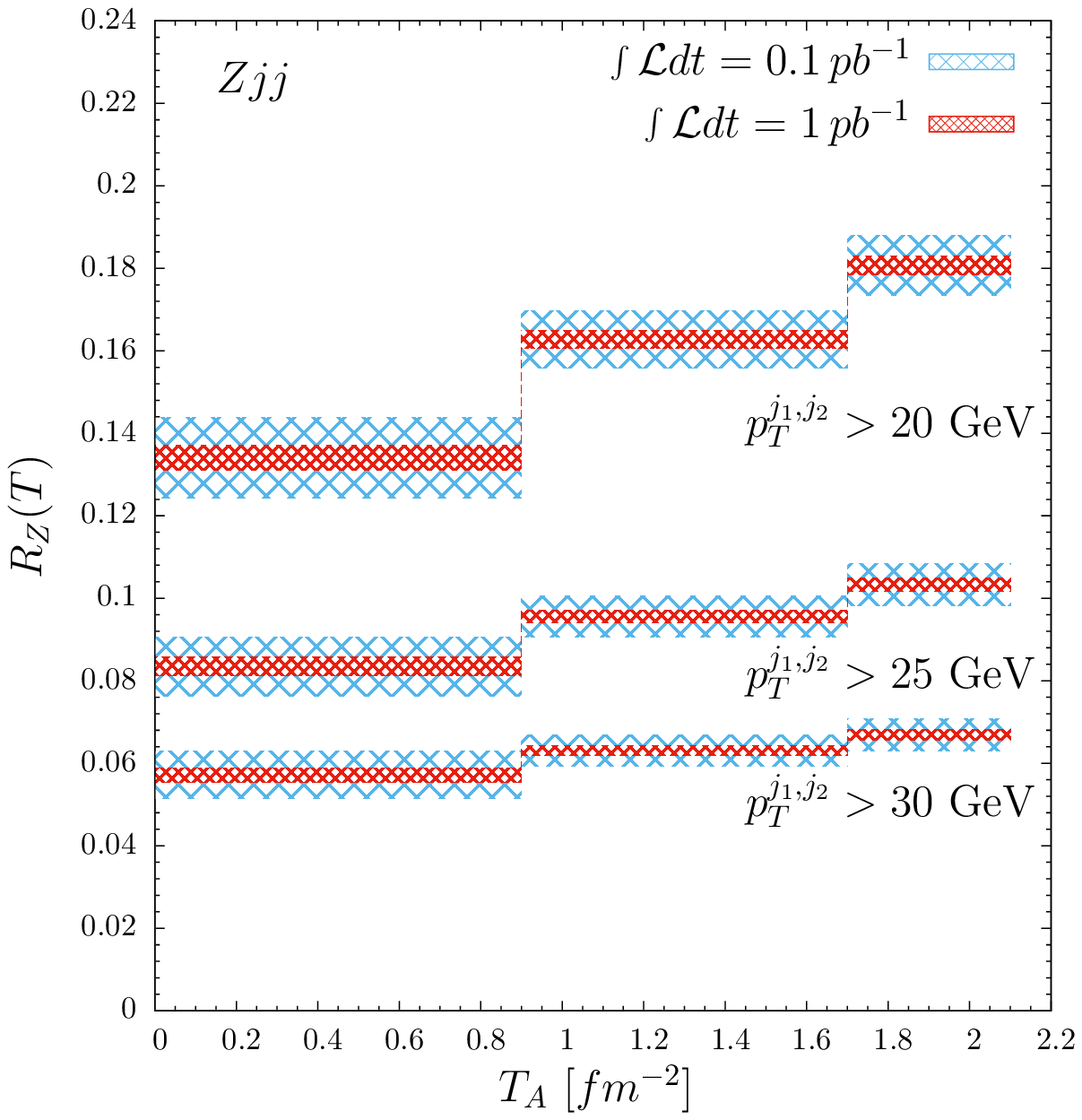}
\caption{\textsl{The ratio of differential cross section $Zjj$ over $Z$ as a function of $T$ (left) and integrated in bins of $T$ (right) assuming two luminosity scenarios for different pt-cuts.}}
\label{plot:rz}
\end{figure}

We present in the left panel of Fig.~(\ref{plot:NevZjj}) the various contributions to the $Zjj$ cross section as a function of the impact parameter $B$ of the $pA$ collisions. 
One may notice there the different behaviour at large $B$ of the various contributions. In the right panel of the same plot we present the expected number of events assuming an 
integrated luminosity of $\int \mathcal{L} \mbox{dt}= 0.1 \, \mbox{pb}^{-1}$, a value in line with data recorded in 2016 $pA$ runs, integrated in bins of $B$.
Even within such a worst-case luminosity scenario, the expected number of events 
allows for an analysis to be performed. 
In addition we present in Tab.~(\ref{Nev_zjj_LO_sym}) the expected 
number of events, $N_{ev}$, as a function of $T$.
With these numbers at our disposal we may exploit the different dependence on $T$ of the various DPS and SPS contributions. 
For this purpose we consider the ratio  $R_{Z}$ 
between the total number (DPS+SPS) of $Zjj$ events over those for
$Z$ production as a function of $T_A(B)$:
\begin{equation}
R_{Z}(T) = N_{Zjj} (T)/ N_{Z}(T).
\label{rzjj}
\end{equation}
In such a ratio, $N_{Z}(T)$ is linear in $T_A(B)$, as well as the SPS background and DPS1 mechanisms contributing to $N_{Zjj}$.
Therefore, in the absence of the quadratic DPS2 contribution, 
the ratio would be a constant. Its deviation from such a behaviour will be just due to DPS2 contribution, which will determine the slope of its linear increase.
The resulting distribution is presented in Fig.~(\ref{plot:rz}) 
for different values of jet transverse momenta cut off 
as a function of $T$ (left) and integrated in bins of $T$ (right). The rise of the slope is related to fast rise of the dijet cross sections entering the DPS2 estimation as the cuts on jet transverse momenta are decreased.
By assuming that statistical error follow a Poissonian distribution, the projected error is calculated from the expected number of events. Our error estimate indicates
that the departure from a constant behaviour can be appreciated in the already recorded 2016 data and that lowering the cut on jets transverse momenta increases the significance of the result.

\section{Next to Leading order results}
\label{Sec:NLO}

The leading order analysis of the previous section is indeed useful to gauge
the order of magnitude of the effects we are searching for. However 
precision phenomenology is nowadays achieved by performing, at least, a  next-to-leading order analysis. All the cross sections relevant for the $Zjj$ final state, both in SPS and DPS are known at parton level, at least, to such an accuracy. In this Section we present the results of this improved analysis.
\par  There is however a caveat which must be addressed at this point. As it is well known in the literature~\cite{FR}, dijet cross sections at next-to-leading order are  not reliable when symmetric cut on jet transverse momenta are enforced, despite the observable being infrared safe. Let us stress that this is not the artifact of numerical simulations. The origin of this instability was first investigated and explained in Ref.~\cite{FR}.
 In  a symmetric (or nearly to symmetric) jet cuts configuration,
the cross section is dominated by a nearly back-to-back kinematics 
and the phase space for the emission of a third, real, parton is greatly reduced.  
As a result the contribution coming from soft real radiation is not able to compensate for the large negative contribution of the soft-virtual terms which only populates the back-to-back topology. 
In order to recover the predictivity of the theory in such symmetric jet-cut configuration one has to perform an all order soft gluon resummation~\cite{Banfi:2003jj}. 
\setlength{\extrarowheight}{0.2cm}
\begin{table}[t]
\begin{center}
\begin{tabular}{cccccccc}  \hline  \hline 
 LO & \hspace{0.4cm} DPS1 \hspace{0.4cm} & \hspace{0.4cm} DPS2 \hspace{0.4cm}  
 & \hspace{0.4cm} SPS \hspace{0.4cm} & \hspace{0.4cm} Sum \hspace{0.4cm} 
 & \hspace{0.4cm} $\sigma(Zjj)/\sigma(Z)$ \hspace{0.4cm} &  $f_{DPS1}$  &  \hspace{0.4cm}  $f_{DPS2}$ \\  
$Zjj$ &  [pb] & [pb] & [pb] & [pb] &  & & \\ \hline
$p_T^{j_1,j_2}>30,20$ GeV &  621 &  1632 & 14135 & 16388 & 0.101  &  0.038 &   \hspace{0.4cm}   0.099  \\
$p_T^{j_1,j_2}>35,25$ GeV &  336 &  885  & 10001 & 11223 & 0.069  &  0.030 &   \hspace{0.4cm}   0.079  \\   
$p_T^{j_1,j_2}>40,30$ GeV &  195 &  515  & 7394  & 8105  & 0.050  &  0.024 &   \hspace{0.4cm}   0.064  \\
\hline
NLO & \hspace{0.4cm} DPS1 \hspace{0.4cm} & \hspace{0.4cm} DPS2 \hspace{0.4cm}  
 & \hspace{0.4cm} SPS \hspace{0.4cm} & \hspace{0.4cm} Sum \hspace{0.4cm} 
 & \hspace{0.4cm} $\sigma(Zjj)/\sigma(Z)$ \hspace{0.4cm} &  $f_{DPS1}$  &  \hspace{0.4cm}  $f_{DPS2}$ \\  
$Zjj$ &  [pb] & [pb] & [pb] & [pb] &  & & \\ \hline
$p_T^{j_1,j_2}>30,20$ GeV &  1286 &  3382 & 14754 & 19423 & 0.094  &  0.066 &   \hspace{0.4cm}   0.174   \\
$p_T^{j_1,j_2}>35,25$ GeV &  684  &  1800 & 10575 & 13059 & 0.063  &  0.052 &   \hspace{0.4cm}   0.138   \\   
$p_T^{j_1,j_2}>40,30$ GeV &  385  &  1013 & 7714  & 9112  & 0.044  &  0.042 &   \hspace{0.4cm}   0.111   \\
\hline
\end{tabular}
\caption{\textsl{Predictions for $Zjj$ DPS and SPS cross sections in $pA$ collisions in fiducial phase space, for asymmetric cuts on jets transverse momenta evaluated at leading (upper) and next to leading accuracy (bottom).}}
\label{Zjj:NLOcs}
\end{center}
\end{table}

\par  A more practical approach, which will be used here, is to use asymmetric cuts on jet transverse momenta, as originally proposed in Ref.~\cite{FR}. 
We performed a series of NLO dijet simulations using \texttt{NLOjet++} \cite{NLOjet}  and 
progressively increasing the imbalance on the jet $p_T$-cuts.  
We found that imbalances larger than 10 GeV allow us to obtain pretty reliable results, while smaller imbalances lead progressively to the exposure of the large and negative contribution 
of the soft-virtual term on the cross section in the lowest $p_T$-bin, as it was explained above.
We therefore choose 10 GeV as our default imbalance in all simulations
The selection  of cuts in the asymmetric configuration is chosen to be 
$p_T^{j_1,j_2}>$30,20 GeV, $p_T^{j_1,j_2}>$35,25 GeV and $p_T^{j_1,j_2}>$40,30 GeV
where $j_1$ represents the leading jet and $j_2$ the subleading one. Jets, again, are clustered at parton level according to anti-$k_t$ jet algorithm with jet radius $R=0.7$.
\par  As it will be apparent from the results presented below, these additional 
cuts severely reduce the number of dijet events contributing to the DPS 
signal and therefore a full analysis must be repeated. 
\par The $Z$ and SPS $Zjj$ cross sections are both evaluted with MCFM 9.0 \cite{MCFM}
at next to leading order accuracy by using NLO CTEQ6M~\cite{Pumplin:2002vw} parton distributions
supplemented with nuclear effects from Ref.~\cite{EKS98}. 
The renormalization and factorization scales for all the relevant processes are fixed as in Sec. \ref{Sec:LO}.

\par Let us   briefly review the impact of NLO corrections to the various cross sections. The $Z$ fiducial cross section is enhanced by a factor 1.3 with respect to the LO one. The dijet cross section is enhanced by a factor of 1.6 whereas the SPS 
$Zjj$ background is only augmented by 4\%, in line with results of Ref.~\cite{Anger:2017nkq}. 
All the analysis is repeated also at leading order, in order to have an indication on the impact of NLO correction on the results. 
\par In Tab.~\ref{Zjj:NLOcs} we report the LO and NLO results for 
all the relevant cross sections within different jet cut configurations.

\setlength{\extrarowheight}{0.2cm}
\begin{table}[t]
\begin{center}
\begin{tabular}{cccccc}  \hline  \hline 
 NLO && \hspace{0.4cm}  $p_T^{j_1,j_2}>30,20$ GeV \hspace{0.4cm}  & \hspace{0.4cm}  $p_T^{j_1,j_2}>35,25$ GeV \hspace{0.4cm}  & \hspace{0.4cm}  $p_T^{j_1,j_2}>40,30$ GeV \hspace{0.4cm}  & \\ \hline
 $T_{min}$  & $T_{max}$ &  $N^{Zjj}_{ev}$ &    $N^{Zjj}_{ev}$ &   $N^{Zjj}_{ev}$ &  $N^{Z}_{ev}$ \\ \hline
  0.00 &  1.60 &  910  &  618  &  435  &  10198 \\
  1.60 &  2.10 &  1032 &  687  &  476  &  10471 \\ \hline
\end{tabular}
\caption{\textsl{Number of expected $Zjj$ and $Z$ events assuming  $\int \mathcal{L} \mbox{dt}= 0.1 \, \mbox{pb}^{-1}$,  integrated in bins of $T_A$
with $Z$ decaying into opposite sign electrons and muons. The Number of $Zjj$ events is reported for three different cuts of jet transverse momenta.
Cross sections are evaluated to next to leading order accuracy.}}
\label{Nev_zjj}
\end{center}
\end{table}

\begin{figure}[t]
\includegraphics[scale=0.6]{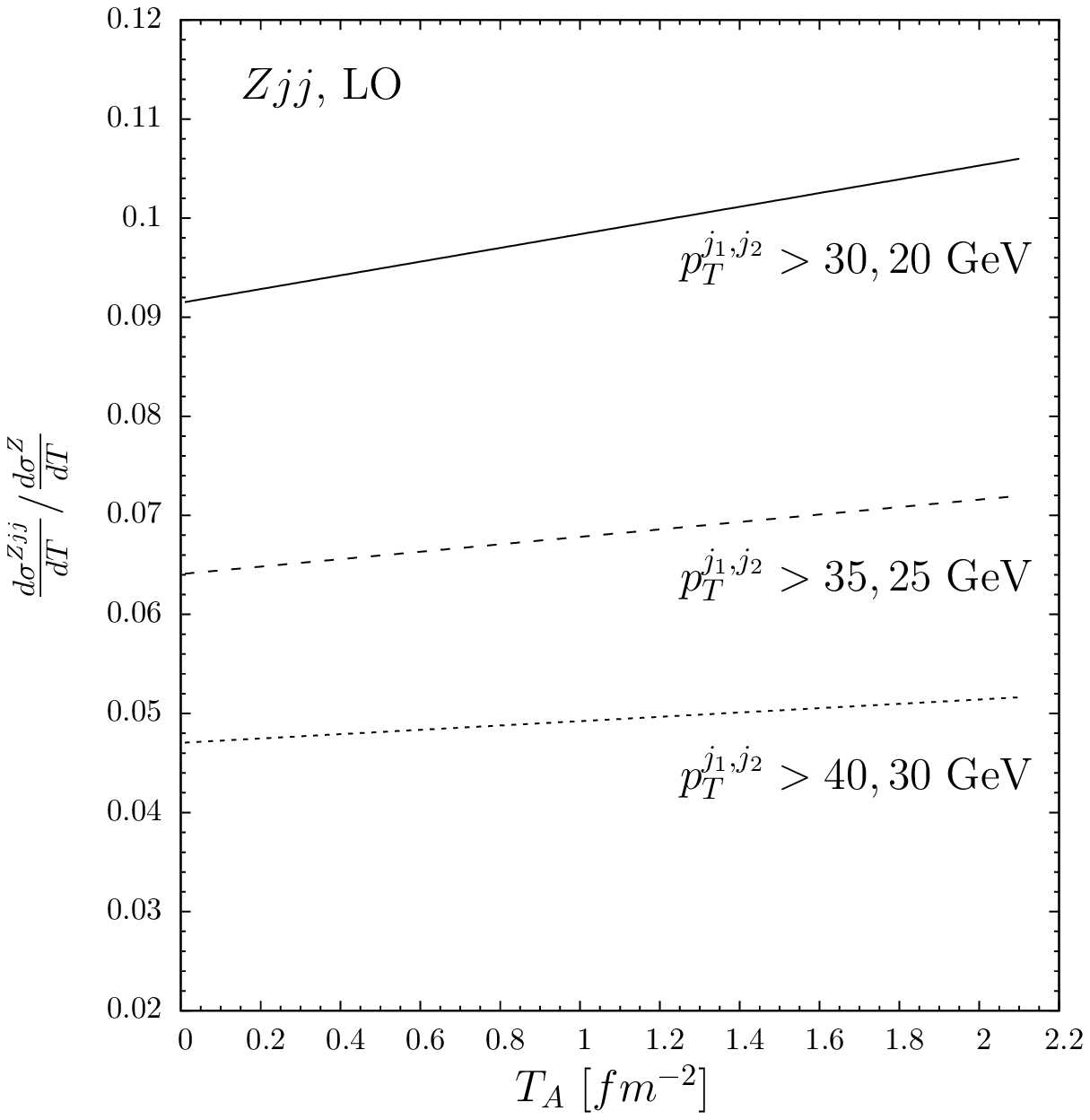}
\includegraphics[scale=0.6]{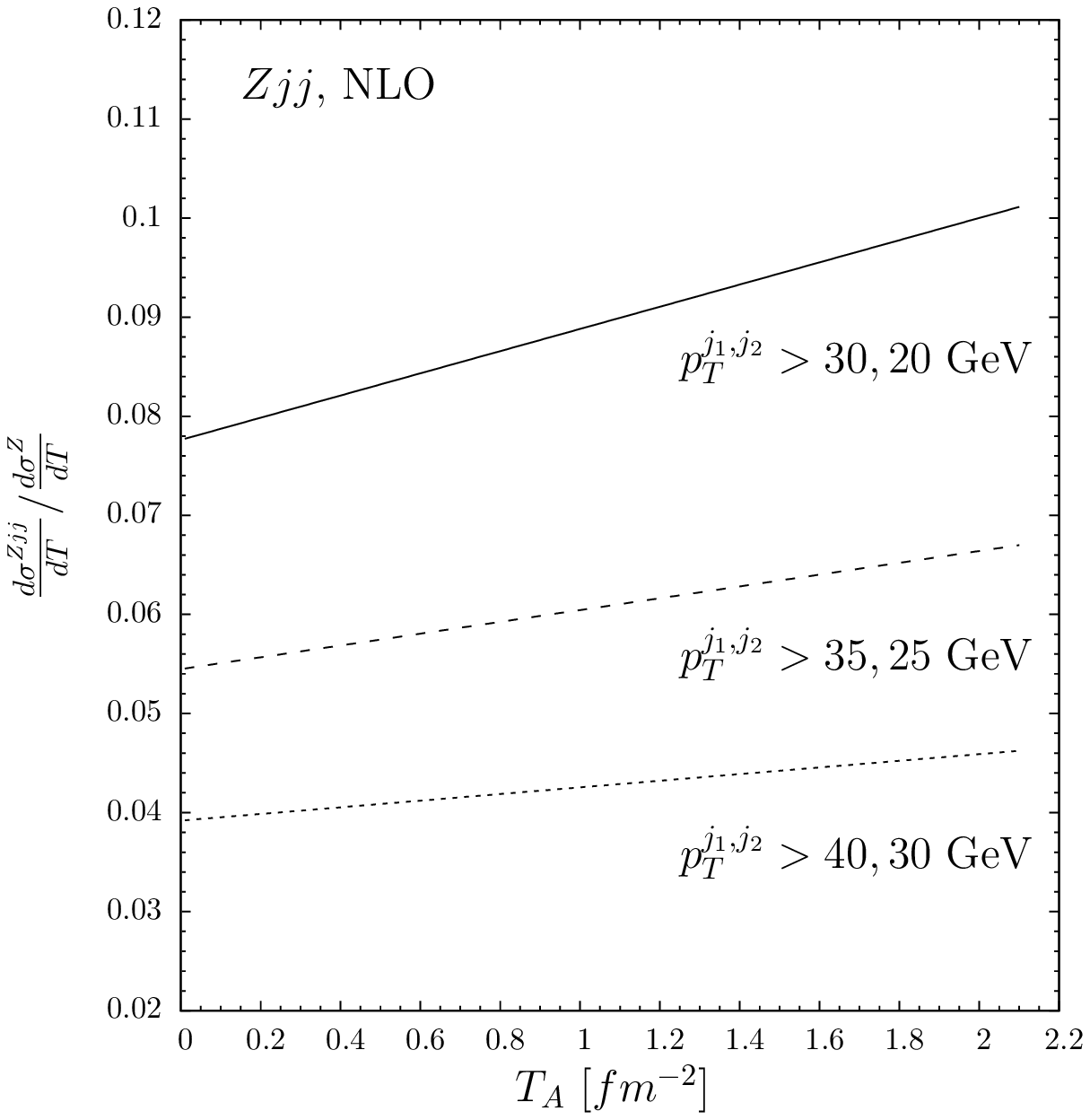}
\caption{\textsl{The ratio of $Zjj$ differential cross section over $Z$ as a function of $T$ evaluated to LO (left) and NLO accuracy (right) . The ratio is evaluated for three asymmetric pt-cuts configurations.}}
\label{plot:NLOrz_line}
\end{figure}

\begin{figure}[t]
\includegraphics[scale=0.6]{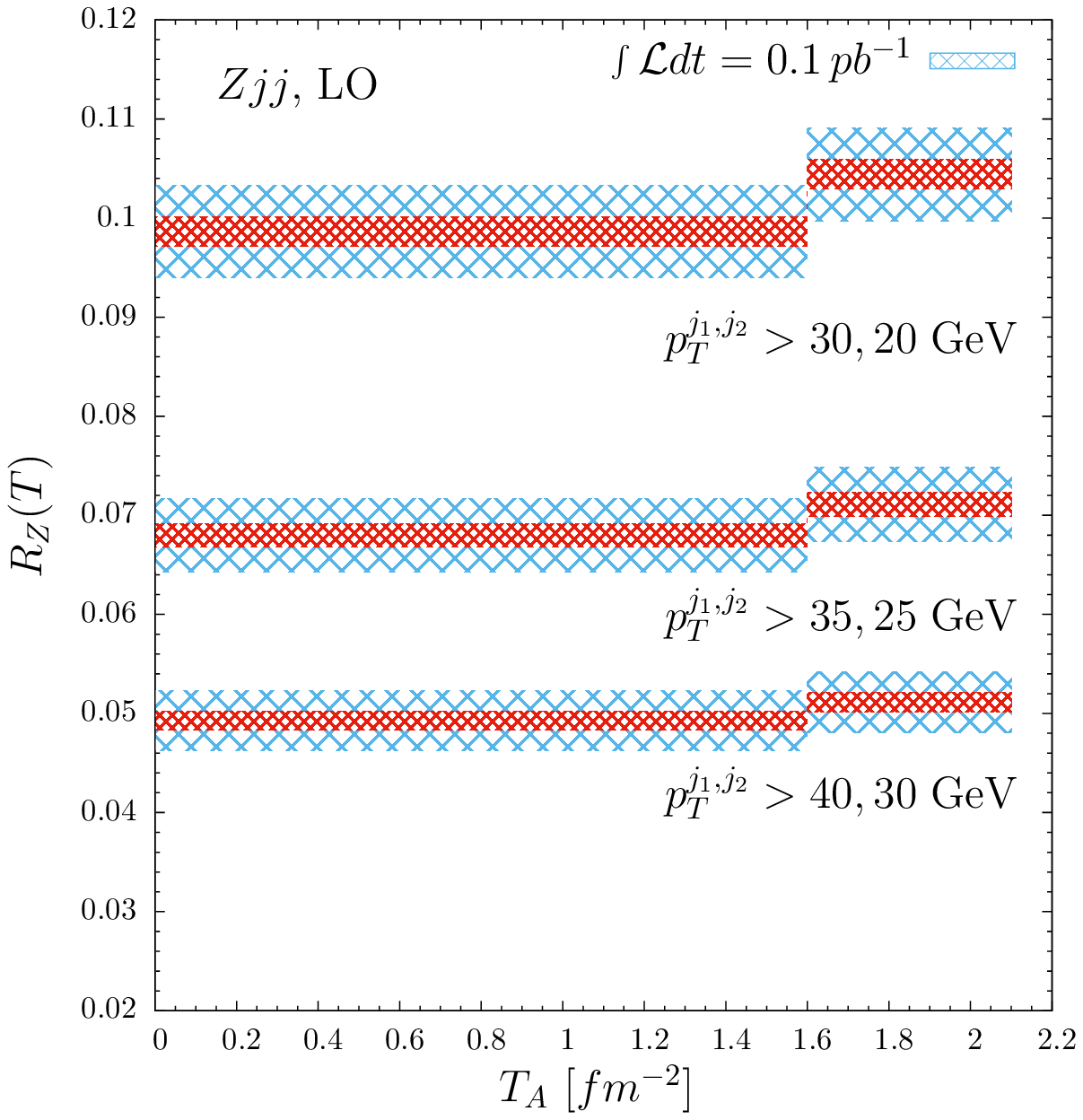}
\includegraphics[scale=0.6]{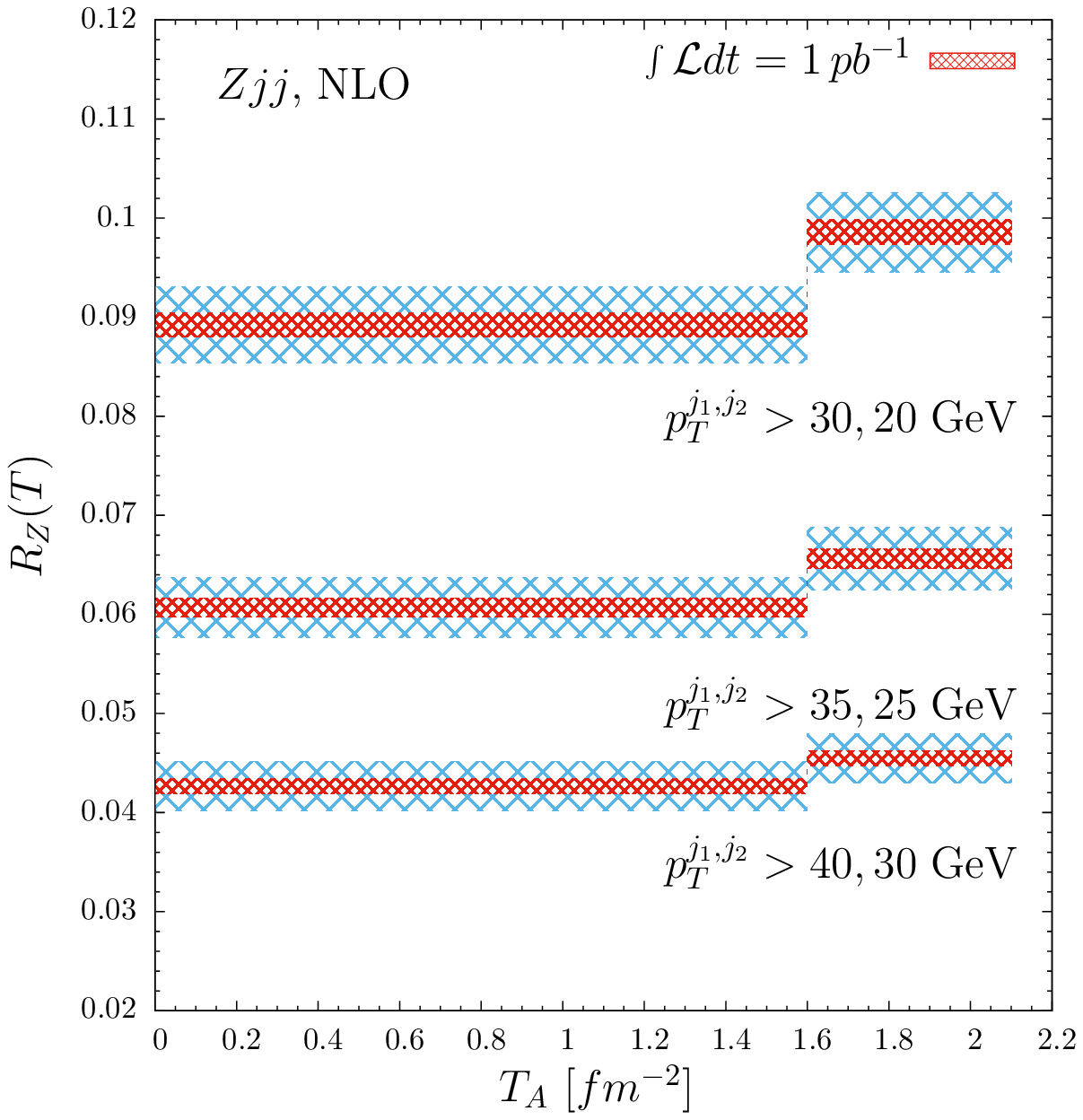}
\caption{\textsl{The ratio $R_Z$ in bins of $T$ evaluated to LO (left) and NLO (right) accuracy. The outer (inner) band stems for projected statistical error with $\int \mathcal{L} \mbox{dt}= 0.1 (1)\, \mbox{pb}^{-1}$. The ratio is evaluated for three asymmetric pt-cuts
configurations.}}
\label{plot:NLOrz}
\end{figure}

In Tab.~\ref{Nev_zjj} we present the NLO results for 
the expected number of events by assuming $\int \mathcal{L} \mbox{dt}= 0.1 \, \mbox{pb}^{-1}$, integrated in bins of $T$.
In Fig.~(\ref{plot:NLOrz_line}) we present
the ratio of the $Zjj$ cross section over $Z$ as a function of
$T$ at leading and next-to-leading order accuracy for different 
cuts on jet transverse momenta. 
Finally, in Fig.~(\ref{plot:NLOrz})
we present, for two luminosity scenarios ($\int \mathcal{L} \mbox{dt}= 0.1,1\, \mbox{pb}^{-1}$),  
the ratio $R_Z$ evaluated at LO and NLO.
\par The comparison of these plots allows to estimate the impact of NLO corrections. 
The value of the ratio is slightly reduced by moving from LO to NLO, an effect related to increased $Z$ cross section appearing in the denominator, given that the dominant SPS $Zjj$ contribution is only slightly augmented at NLO. On the other hand the slope of the distributions is steeper
at NLO since the dijet cross sections has a large K-factor, enhancing its contribution over its LO estimates. 
Therefore NLO corrections act in making 
more evident the non constant behaviour of the $R_Z$ ratio. 
In addition one may notice that also the statistical error associated with the predictions is slightly reduced from going LO to NLO (for the same cuts), given the larger number of inclusive $Z$ events at NLO. 
For asymmetric cuts configuration, as expected, the statistics is worse than in the symmetric set up, but still a non constant behaviour as a function of $T(B)$ can be observed with the lowest $p_T$-cut threshold. Future $pA$ runs 
accumulating more than $1 \, \mbox{pb}^{-1}$ will allow to observe a non-constant behaviour in all cut configurations. Note that, due to the significant decrease of statistics, we were forced to present the results for two bins and not for three as for the symmetric-cuts LO case.
\par  Finally we make a comment on the symmetric cuts configuration at NLO. If we assume that NLO corrections in that case follows the pattern of the asymmetric case, one may expect a steeper slope of the ratio $R_Z$ even in those configurations, therefore giving an enhanced sensitivity to the DPS2 signal, on top of the larger statistics attainable with the symmetric cut configuration.

\section{Conclusions}
\label{Sec:Conclusions}
\par We have calculated 
cross sections for $Zjj$ final state in $pA$ collisions at the LHC with the aim of studying the so called DPS2 contribution to the cross section. We have shown that for symmetric 
cuts the separation of DPS2 contribution can be made already with data recorded in 2016 in dedicated $pA$ runs and will definitely improve for future runs at the LHC. 
\par In addition we made the first step in the study of  the NLO contributions.  In this case we had to choose  asymmetric cuts configuration in order to deal with well known instabilities of NLO dijet cross sections. To NLO accuracy the sensitivity
to the DPS2 contribution is increased due to the large 
K-factor of dijet cross sections in going from LO to NLO. 
However the use of asymmetric cuts has the obvious disadvantage of reducing statistics and the significance of the results.
In order to recover the statistics granted by symmetric cuts
configuration, and to perform the NLO analysis for the symmetric cuts, one would need more refined dijet predictions, \textsl{i.e.} including soft gluon resummation, whose inclusion is, however, beyond the scope of the current paper.
Despite this observation, within the different sets  of cuts and at next-to-leading order accuracy, the DPS2 contribution
has large enough cross sections to allow its determination already with data recorded in 2016 in dedicated $pA$ runs 
if the jet cut are retained as low as possible and it could be 
definitely observed for future runs at the LHC with higher integrated luminosity.

\section*{Acknowledgement}
We thank A. Milov and M. Strikman for very useful discussions. We also warmly thank
Daniel Britzger and Zoltan Nagy for their support with NLOjet++ and Tobias Neumann with MCFM. 
The work was supported by Israel Science  Foundation  under  the  grant  2025311.


  \end{document}